An improved technique for measuring plasma density to high frequencies on the Parker Solar Probe

by F.S. Mozer[1], S.D. Bale[1], P.J. Kellogg[2], D. Larson[1], R. Livi[1], and O. Romeo[1]
1. Space Sciences Laboratory, University of California, Berkeley, USA
2. University of Minnesota, Minneapolis, Mn, USA

**ABSTRACT**

The correlation between the plasma density measured in space and the surface potential of an electrically conducting satellite body with biased electric field detectors has been recognized and used to provide density proxies. However, for Parker Solar Probe, this correlation has not produced quantitative density estimates over extended periods of time because it depends on the energy dependent exponential variation of the photoemission spectrum, the electron temperature, the ratio of the biased surface area to the conducting spacecraft surface area, the spacecraft secondary or thermal emission, the spacecraft distance from the Sun, etc. In this paper the density as a function of time and frequency to frequencies as high as the electron gyrofrequency is determined through least squares fits of a function of the spacecraft potential to the plasma density measured on the Parker Solar Probe. This function allows correction for the many effects on the spacecraft potential other than that due to the plasma density. Some examples of plasma density obtained from this procedure are presented.

**INTRODUCTION**

Since electric field antennas were first biased to make double probe electric field measurements in space plasmas [Mozer and Bruston, 1967] it has been recognized that the potential of the spacecraft body with respect to infinity, $v_s$, is related to the ambient plasma density [Celnikier et al, 1983; Knott et al, 1984; Marsch et al, 1990; Pedersen, 1995; Kellogg et al, 2005; Hnat et al, 2005; Chen et al, 2012]. That the spacecraft potential is related to the plasma density follows from the requirement that the net current to the spacecraft body be zero in equilibrium, or

$$I_{pe} - I_e + I_0 + I_1 = 0 \qquad (1)$$

where
  $I_{pe}$ is the part of the photoemission current that escapes to infinity

$I_e$ is the plasma electron thermal current to the body
$I_0$ is the bias current that is applied from the spacecraft to the antennas
$I_1$ = the current due to all other sources, including ions, secondary emission, thermal emission, etc.

When a spacecraft's photoemission exceeds its random electron thermal current collection and all other currents are smaller, the spacecraft charges positive with respect to infinity until the photocurrent that escapes to infinity, $i_{pe}\exp(-v_s/v_o)$, equals the plasma thermal current, where $v_o$ is the energy dependent e-folding energy of the photoemission spectrum. Because the random plasma thermal current is proportional to the plasma density, the requirement of no net current to the body produces a relationship between the potential of the spacecraft and the local plasma density. This relationship is complex because the photoemission may not always exceed the plasma electron current, the photoemission spectrum is not exactly an exponential [Scudder et al, 2000], secondary and thermal emission from the spacecraft surface may be important on the Parker Solar Probe, etc. It is assumed that these effects vary more slowly than the density such that calibrations of the potential-density relation on a short time scale produces density measurements from the spacecraft potential. For this work, the spacecraft potential was obtained from the Parker Solar Probe Fields instrument[Bale et al, 2016; Malaspina, 2016; Mozer et al, 2020], the electron density was measured by the Solar Probe ANalyzer (SPAN)-Electron sensor [Whittlesey et al, 2020] as described by Halekas et al, [2020], and the plasma density was obtained from the quasi-thermal noise spectrum of the electric antenna near the plasma frequency, as described by Meyer-Vernet and Perche [1989].

**THE RELATIONSHIP BETWEEN THE SPACECRAFT POTENTIAL AND PLASMA DENSITY**

In the above equation (1)

$I_{pe}$ = current due to escaping photoelectrons
   = $i_o A_s (R_o/R)^2 \exp(-v_s/v_o)$
and
   $A_s$ = area of the sunlit spacecraft surface
   $R$ = distance from the Sun
   $v_s$ = potential of the spacecraft with respect to infinity, which is assumed to be positive because the photoelectron flux exceeds all other currents.
   $v_o$ = e-folding energy of the assumed exponential photoelectron distribution.

$I_e$ = plasma electron thermal current
   = $neA_c[kT/2\Pi m]^{0.5}(1+ev_s/kT)$

And
   n = electron density
   e = electron charge
   $A_c$ = spacecraft current collecting area
   T = electron temperature
   m = electron mass
   $(1+ev_s/kT)$ = focusing factor [Mott-Smith and Langmuir,1926]

   $I_0 = i_0 A_A$ = the bias current that is applied from the spacecraft to the antennas, where $i_0$ is a unit bias current and $A_A$ is the area of the antennas.

   $I_1$ = the current due to all other sources, including ions, secondary emission, thermionic emission, etc.

Solving equation (1) after lumping all constants into $K_1$, $K_2$ and $K_3$ gives

$-v_s = K_1 \ln[nR^2(A_C/A_s)T^{0.5}] + K_2 \ln(1+ev_s/kT) + K_3 \ln[1+(i_0 A_A + I_1)/nA_c T^{0.5}(1+ev_s/kT)]$

The measured spacecraft potential is $V_S = v_A - v_s$, where $v_A$ is the potential of the antennas relative to infinity. Because the antennas are biased to be near zero potential, $v_A$ is small but variable, so it is incorporated into the quantity B in equation (2) and $-v_s = V_s$. Because $ev_s/kT$ is much less than one, the $K_2$ term is small. Because $A_A/A_c$, the antenna area divided by the spacecraft area, is much less than one, and because the neglected current, $I_1$, may be small, the $K_3$ term is also small. The result is thus

   $V_S = A*\ln[nR^2(A_C/A_s)T^{0.5}] + B$                                   (2)

where the constant, B, includes the terms associated with $v_A$, $K_2$ and $K_3$.

On spinning spacecraft, the spacecraft total area divided by the sunlit area, $A_C/A_S$, varies with the spacecraft orientation relative to the Sun, so the above equation does not reproduce the electron density from the spacecraft potential at frequencies greater than the spin frequency unless this factor is accounted for. However, the Parker Solar Probe spacecraft keeps the same surface pointed towards the Sun throughout each perigee pass, so the logarithm of the area ratio is constant and can be included as part of the constant, B. This gives

   $V_S = A*\ln[nR^2 T^{0.5}] + B$                                           (3)

Equation (3) may be validated by comparing $V_S$ with $\ln[nR^2T^{0.5}]$, as is done in Figure 1 where the two quantities are shown to be related. Also illustrated in figure 1 are the core electron perpendicular temperature and the spacecraft location, parameters that enter into the logarithmic function.

In principle, A and B should be nearly constant. However, they are not constant because their included effects are not truly constant or small. For example, the secondary emission from the spacecraft body may be significant and variable, the photoelectron spectrum deviates from an exponential, the area of the spacecraft body is not infinitely large compared to the area of the antennas, the photoemission of the heat shield may vary with temperature, the potential of the antennas with respect to infinity may not be small, etc. These shortcomings are overcome partially by incorporating these error terms into the constants A and B and doing least squares fits to find A and B on time scales during which the error terms are roughly constant. Thus, A and B are determined from four minute least squares fits of equation (2), using 30 second average measurements of spacecraft potential, electron temperature, density, and the spacecraft location. Following this step, the two constants are coupled with the highest time resolution measurements of the spacecraft potential to produce high time resolution plasma density from the spacecraft potential. The 30 second averaging time scale was selected from the desire to average many samples of the noisy <1 Hz plasma data into a single analysis data point. The four minute least squares fitting time scale was selected from the desire to have eight data points in each least squares fit.

Figure 2 displays the results of such a least squares fit. In panel 2A, the red curve is the plasma density determined from the quasi thermal noise spectrum and the density determined from the higher time and frequency resolution spacecraft potential measurement is the black curve. The two curves are identical except for the higher frequency variations determined from the spacecraft potential. The bottom panel of Figure 2 displays these higher frequency variations after the data is high-pass filtered at 0.5 Hz. The density fluctuations, appearing as blobs in time in panel 2B, have amplitudes as large as 500 cm$^{-3}$ when the background density is less than 7000 cm$^{-3}$, so large, low frequency, density fluctuations are present.

The one hour averaged coefficients A and B of the least squares fits are shown for the two days of interest in figure 3. The coefficient, A, is small and positive as is expected for the case of the photoemission exceeding the plasma thermal current.

The upper frequency limit for the validity of the procedure for obtaining density from the spacecraft potential depends on the time constant of the spacecraft potential response to an abrupt change of density. If the density increases, the positive spacecraft potential will decrease due to the collection of more plasma electrons, until current balance is achieved by the escape to infinity of more photoelectrons from the decreasing spacecraft potential. A one-meter radius spherical spacecraft, having a capacitance of $10^{-10}$ farads, will charge by one volt under typical plasma density and temperature conditions in less than one microsecond. Thus, the procedure for obtaining the plasma density from the spacecraft potential should be valid to frequencies much greater than the 10 kHz frequency illustrated in following figures.

A conceivable error in the determination of the plasma density from the spacecraft potential is that the photoelectron spectra are not exponentials over a wide energy range. However, they may be approximated as exponentials over the confined energy and time interval of each least squares fit. In this way, variations of the spectra are accommodated as variations of the least squares coefficients.

As an example of the density computed in the above manner, figure 4 gives electric field (panel 4A) and density (panel 4B) spectra during two intervals, one with and one without waves above a few Hz. During the wave intervals (the black curves), there was electric field and density fluctuation power at 3-10 Hz and again at 100-1000 Hz. Waves at this pair of frequencies are the triggered ion acoustic waves or TIAW [Mozer et al, 2021a, 2021b] that consist of several hundred Hz electrostatic waves that appear in bursts that are phase locked to the low frequency ion acoustic-like waves. During the time interval without such waves (the red curves), the main feature of the spectra was the existence of large electric field power below 10 Hz with no corresponding power in the density fluctuations. This is because these low frequency signals are from electromagnetic waves that were not present during the other interval. This result suggests the possibility that the electrostatic, triggered ion acoustic waves are anti-correlated with low frequency electromagnetic waves.

Figure 5 provides further information on the triggered ion acoustic waves of figure 4 during three time intervals of 10 minutes (panels 5A, 5B, and 5C), one second (panels 5D, 5E, and 5F), and 0.1 seconds (panels 5G, 5H, and 5I). During each time interval, the top plots (panels 5A, 5D, and 5G) give the >100 Hz high pass filtered electric field, the middle plots (panels 5B, 5E, and 5H) give the >0.5 Hz plasma density and the bottom plots (panels 5C, 5F, and 5I) give the >100 Hz plasma density. The bursty electric field was as large as 50 mV/m and the density peaks of 60 cm$^{-3}$ were about 2.5 percent of the background density. This suggests that the pressure associated with density fluctuations may be related to the core electron heating produced by these waves [Mozer et al, 2021b]. These are the first density fluctuation spectra measured at frequencies as high as 10,000 Hz and their correlation with the electric field fluctuations lends credence to the analysis that produced them.

**ACKNOWLEDGEMENTS**

This work was supported by NASA contract NNN06AA01C. The authors acknowledge the extraordinary contributions of the Parker Solar Probe spacecraft engineering team at the Applied Physics Laboratory at Johns Hopkins University. The FIELDS experiment on the Parker Solar Probe was designed and developed under NASA contract NNN06AA01C. Our sincere thanks to P. Harvey, K. Goetz, and M. Pulupa for managing the spacecraft commanding, data processing, and data analysis, which has become a heavy load thanks to the complexity of the instruments and the orbit. We also acknowledge the SWEAP team for providing electron temperature data.

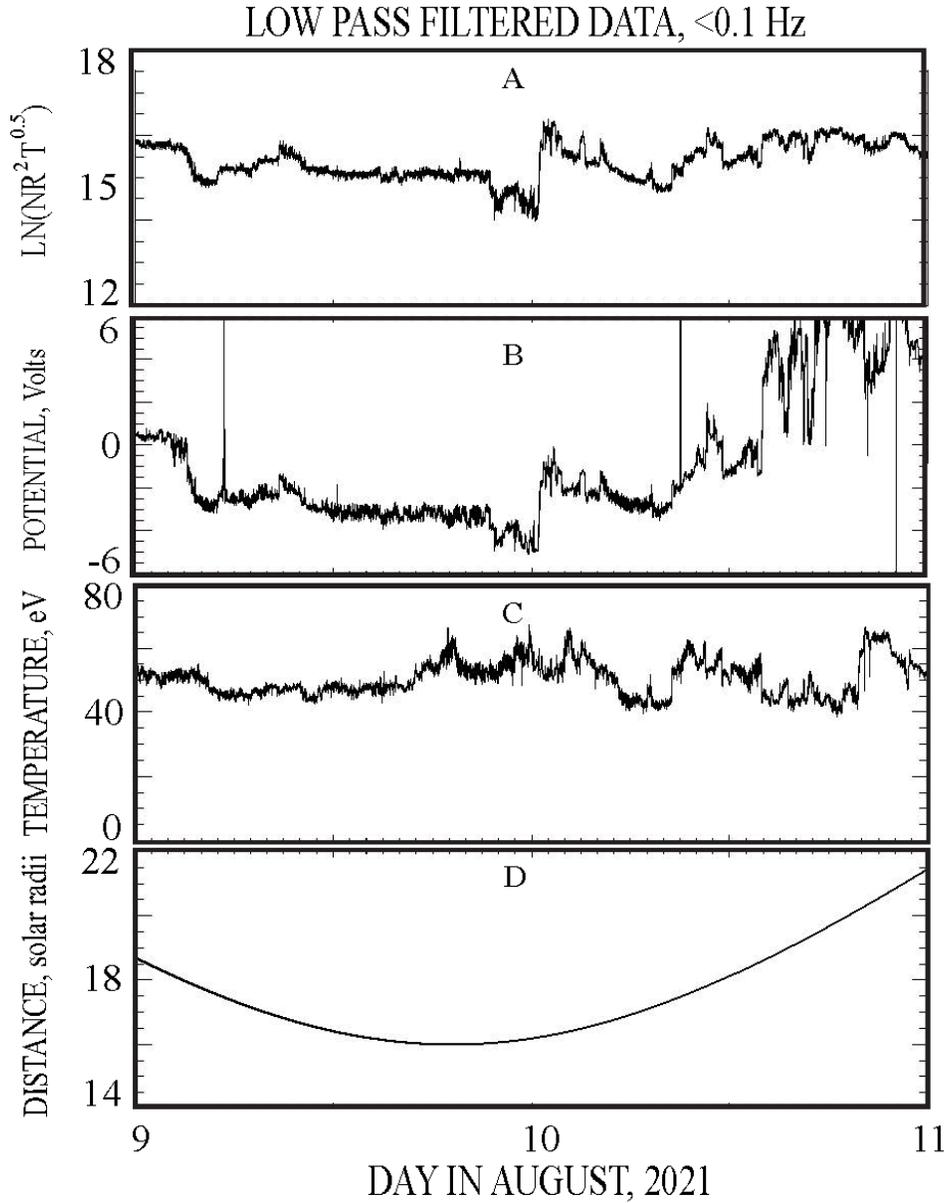

Figure 1. Comparison of $\ln[nR^2T^{0.5}]$ in panel A with the spacecraft potential in panel B. The large vertical lines in panel B are due to dust. The core electron perpendicular temperature and the spacecraft location are given in panels C and D. The general correlation between the quantities in panels A and B and the detailed differences in the amplitudes of their changes show that equation (2) can provide a good fit of the two data sets.

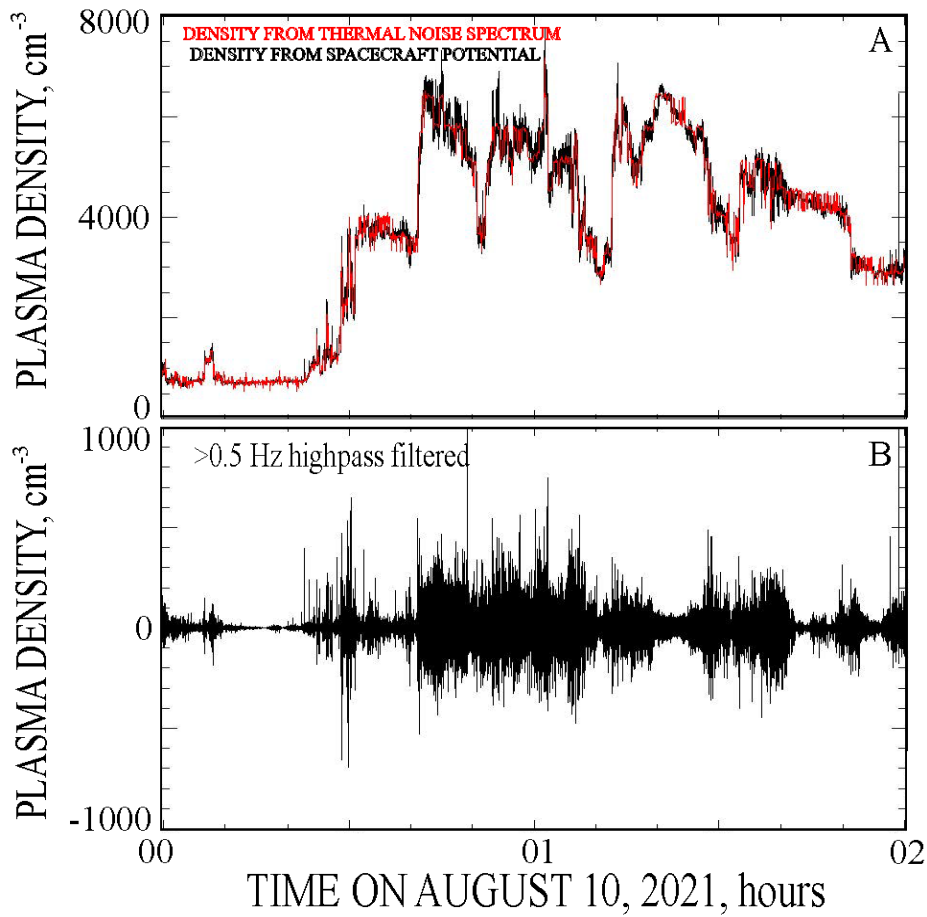

Figure 2. Panel A provides a two-hour comparison between the plasma density determined from the thermal noise spectrum (red) and the density obtained from the spacecraft potential (black). The black curve has a higher frequency response. Its >0.5 Hz high pass filtered data, in panel B, shows that the density fluctuations occurred in bursts and that the low frequency density fluctuations were sometimes as great as 10% of the average density.

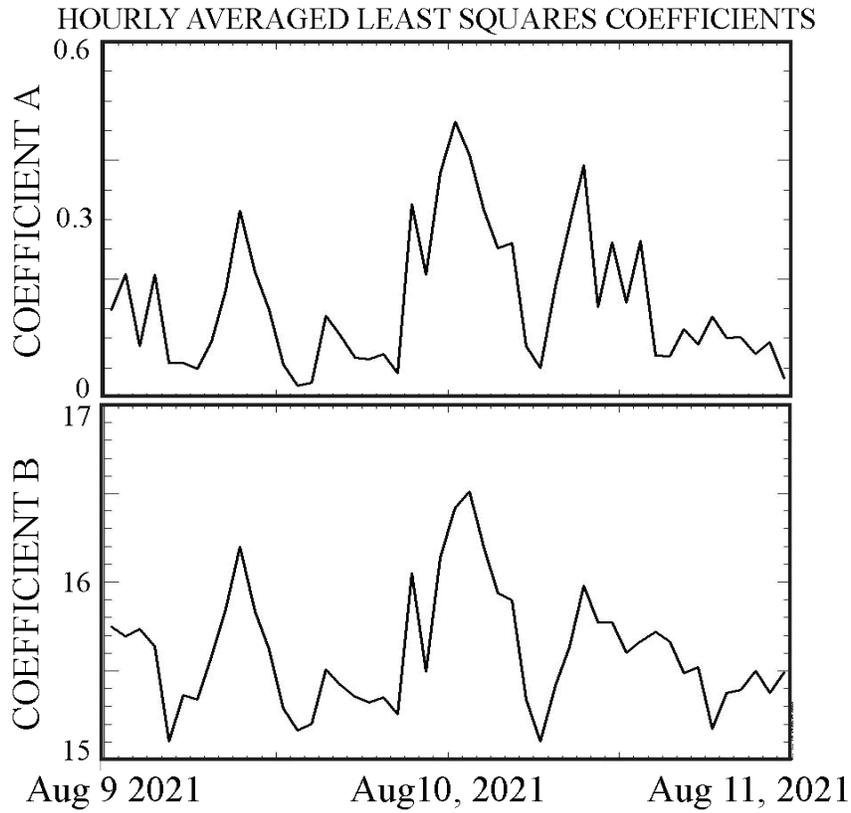

Figure 3. Coefficients A and B obtained from the least squares fitting of equation 2. Note that coefficient A is greater than zero, which it must be if the photoemission exceeded the thermal plasma current.

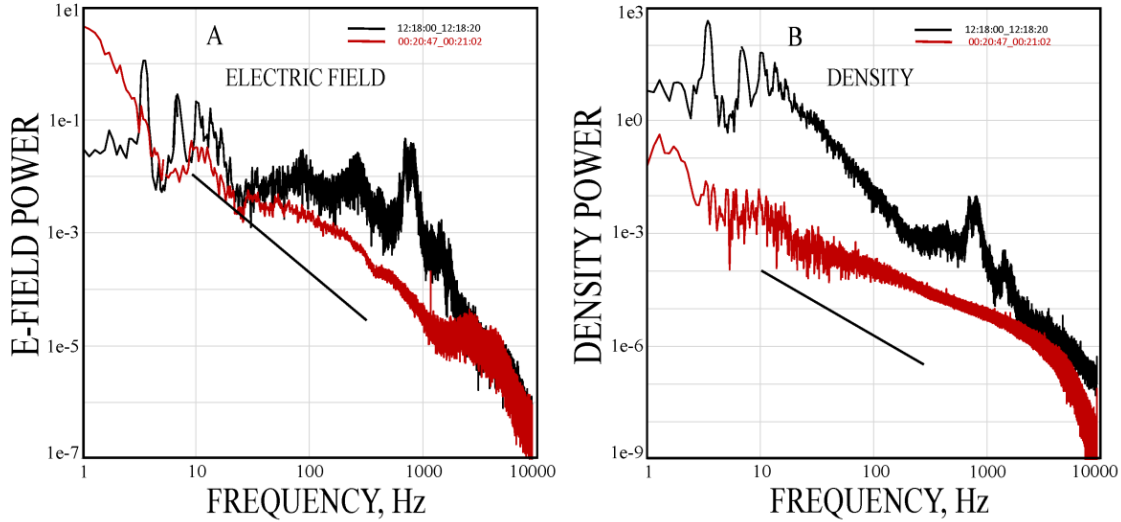

Figure 4. Electric field (panel 3A) and density (panel 3B) spectra during two intervals, one with waves (the black curves) and one without waves above 10 Hz (the red curves). The straight lines have a -5/3 spectral slope. As judged from the presence or absence of density fluctuations, the waves that produced the black curves were mostly electrostatic while the low frequency waves that produced the red curves were mostly electromagnetic.

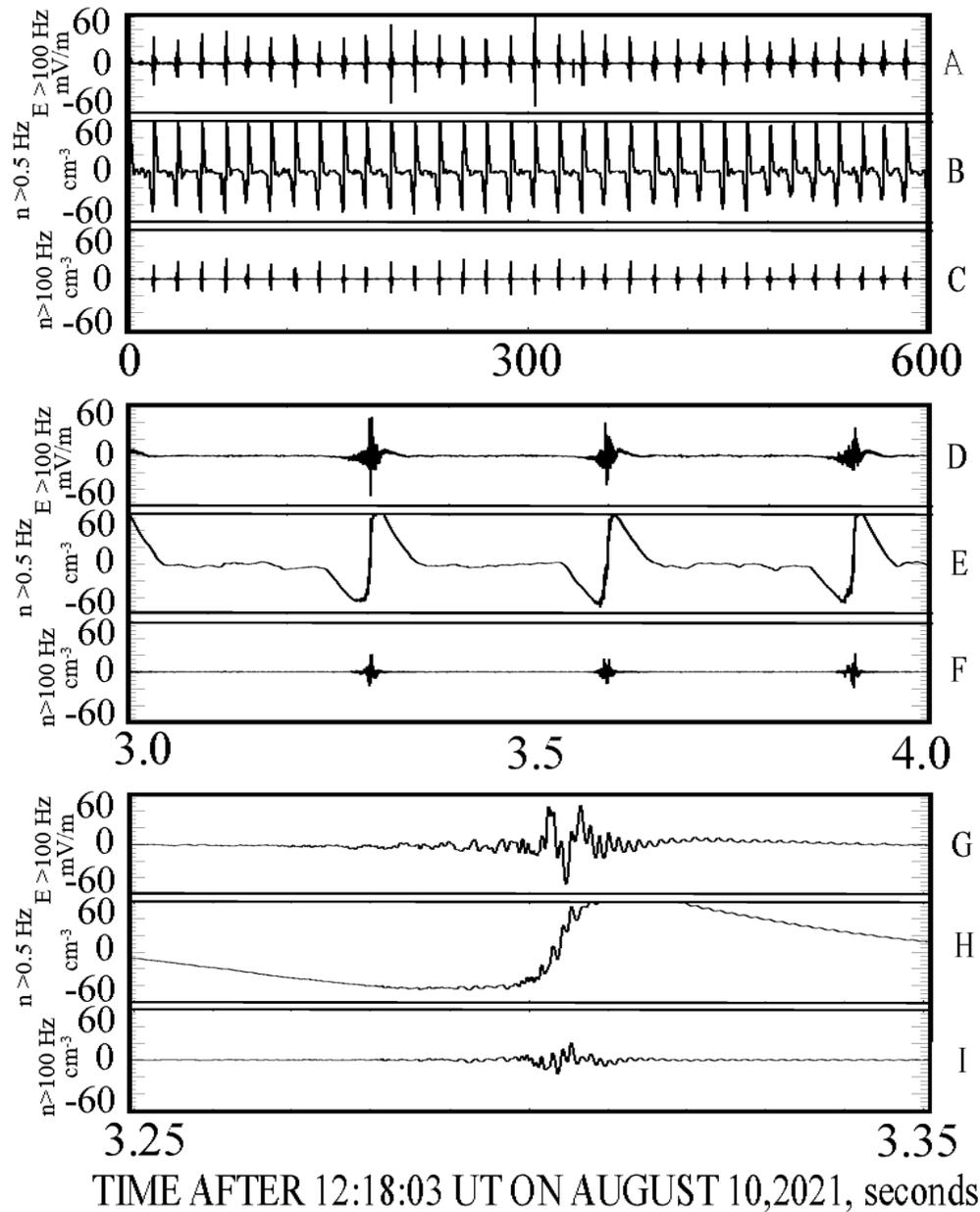

Figure 5. >100 Hz electric field (panels 5A, 5D, and 5G), >0.5 Hz plasma density fluctuations (panels 5B, 5E, and 5H) and >100 Hz plasma density fluctuations (panels 5C, 5F, and 5I) observed during time intervals of 10 minutes, 1 minute and 0.1 seconds, respectively.

12